\documentclass{article}
\usepackage{amsmath,amssymb,epsfig,amsthm,cite}

\def\beq{\begin{eqnarray}}
\def\eeq{\end{eqnarray}}
\def\bsp{\begin{split}}
\def\esp{\end{split}}

\def\ra{\rightarrow}

\newcommand{\mb}[1]{{\mathbb #1}}
\newcommand{\mc}[1]{{\cal #1}}

\newtheorem{thm}{Theorem}

\begin{document}

\title{\textbf{The Asymptotic Behaviour of Tilted Bianchi type VI$_0$
    Universes}} 
\author{\textbf{Sigbj\o rn Hervik}\thanks{S.Hervik@damtp.cam.ac.uk} \\
DAMTP, \\
Centre for Mathematical Sciences,\\
Cambridge University\\
Wilberforce Rd. \\
Cambridge CB3 0WA, UK}
\maketitle
\begin{abstract}
We study the asymptotic behaviour of the Bianchi type VI$_0$
universes with a tilted $\gamma$-law perfect fluid. The late-time attractors are
found for the full 7-dimensional state space and for several
interesting invariant subspaces. In particular, it is found that for the
particular value of the equation of state parameter, $\gamma=6/5$,
there exists a bifurcation line which 
signals a transition of stability between a non-tilted equilibrium
point to  
an extremely tilted equilibrium point. The initial singular regime is
also discussed and we argue that the initial behaviour is chaotic for $\gamma<2$.

\end{abstract}
\section{Introduction} 
In a recent paper we analysed how tilted fluids affect the general
late-time behaviour of cosmological models of Bianchi type
\cite{BHtilted}. As the asymptotic behaviour of all Bianchi models with a
non-tilted $\gamma$-law perfect fluid are determined \cite{DynSys,HHTW,HHWVIII},
the sensitivity of the non-tilted equilibrium points was checked with
regards to inclusion of tilted fluids. In this paper, we will consider one of
the models in full generality; namely the Bianchi type VI$_0$ model
containing a tilted $\gamma$-law perfect fluid. 

The last decades, there have been some investigations of Bianchi
models with a tilted fluid \cite{KingEllis,BN,BS}; in particular, 
type II \cite{HBWII} and type V \cite{Shikin,Collins,HWV}. Apart from
these special Bianchi types, the general behaviour of tilted Bianchi
universes seems to be poorly understood. The work \cite{BHtilted} gave
us some hints  where interesting behaviour may occur, but a
more elaborate analysis is needed in order to understand the more
general behaviour of universes with tilted fluids. More specifically,
in \cite{BHtilted} it was shown that for the Bianchi type VI$_0$ model
there are no stable non-tilted equilibrium points for equation of
state parameter obeying $\gamma>6/5$. The value $\gamma=6/5$ signals
the onset of an instability with respect to tilt implying 
that at late times the peculiar velocities of the fluid is a major
contributor to the cosmological shear. 
The Bianchi type VI$_0$  is not the most general model but is
sufficiently general to 
account for many interesting phenomena. Thus we will assume that the
universe contains a $\gamma$-law perfect fluid ($0<\gamma<2$) which in general can be
tilted. The state space is thus 7-dimensional \cite{DynSys} (compared
to the 8-dimensional state space for the most general models). 

The dynamical systems approach -- which is the method adopted here --
requires that we find all equilibrium points for the system of
equations. These equilibrium points play a special role
in the evolution of the system. Not only do they correspond to exact 
self-similar solutions to Einstein's equations, but  some of them may even
serve as attractors for more general solutions. However, not many 
 exact solutions with tilted fluids are known for the Bianchi models. Some type
II solutions have been found \cite{HewittII}, and Rosquist and Jantzen
found some type VI$_0$ solutions \cite{Rosquist,RJ}. Here, we will
show that the solutions (for $6/5<\gamma<3/2$) found by Rosquist and Jantzen are attractors
in a four-dimensional invariant subspace of the full state
space. However, in the full state space, there always exists an
unstable mode. In the full state space, the future attractor for
$6/5<\gamma$ is an extremely tilted model which is connected via a line
bifurcation at $\gamma=6/5$ to a non-tilted equilibrium point. This
$\gamma=6/5$ bifurcation, which correspond to a one-parameter family
of exact tilted solutions, seems to have been first discussed in a recent
paper by Apostolopoulos \cite{Apo}. However, not much discussion is
devoted to this solution as the solution itself is implicitly given in terms of a
cubic. Here, we have introduced a different parameterisation which
makes it possible to explicitly write down the solution in terms of
the expansion-normalised variables. 

We also discuss the initial singular regime, which is a
lot more
complicated. In fact, there are indications that the initial singular regime
possesses an oscillatory, very likely even a chaotic behaviour
similar to that found in other Bianchi models
\cite{Spokoiny:1981fs,Belinsky:1970ew,jb2,jb3,cb,dem,DH2000,dh2,hbc}. 

The paper is organised as follows. Next, in section \ref{sect2}, we
write down the equations of motion using the orthonormal frame
formalism. In section \ref{sect3} we discuss the equilibrium points
which are important for the late-time asymptotic behaviour. Then,
in section \ref{sect4}, we discuss their stability and present some
general results regarding the asymptotic behaviour of the tilted type
VI$_0$ model. The initial singular regime is discussed in section
\ref{sect5} before we finally summarise our results  in  section \ref{sect6}. 

\section{Equations of motion}
\label{sect2}
The energy-momentum tensor for a tilted perfect fluid is 
\beq
T_{\mu\nu}=(\hat\rho +\hat{p})\hat{u}_{\mu}\hat{u}_{\nu}+\hat pg_{\mu\nu},
\eeq
where $\hat{u}^{\mu}=(\cosh\beta,\sinh\beta c^a)$ is the fluid velocity. The
spatial vector $c^a$ is chosen to be a unit vector in the tangent space of the surfaces of homogeneity; i.e.
$c^ac_a=1$. We will further assume that the fluid obey the barotropic
equation of state,
\beq
\hat p=(\gamma-1)\hat\rho, \quad 0<\gamma<2.
\eeq
In terms of the unit normal vector ${\bf u}={\bf e}_0$ to the
group orbits the energy-momentum tensor takes the imperfect fluid form 
\beq
T_{\mu\nu}=(\rho+p) u_{\mu}u_{\nu}+pg_{\mu\nu}+2q_{(\mu}u_{\nu)}+\pi_{\mu\nu},
\eeq
where 
\beq
\rho&= &(1+\gamma \sinh^2\beta)\hat\rho,\\
p&=&\left(\gamma-1+\frac 13\gamma\sinh^2\beta\right)\hat\rho, \\
q_{a}&=&\gamma \hat\rho\cosh\beta\sinh\beta c_a,\\
\pi_{ab} &=& \gamma\hat\rho\sinh^2\beta\left(c_ac_b-\frac 13 h_{ab}\right).
\eeq

 For the Bianchi cosmologies -- which admit a simply
transitive symmetry group acting on the spatial hypersurfaces -- we can
always write the line-element as 
\[
ds^{2}=-dt^{2}+\delta_{ab}{\mbox{\boldmath${\omega}$}}^{a}{\mbox{\boldmath${%
\omega}$}}^{b},
\]%
where ${\mbox{\boldmath${\omega}$}}^{a}$ is a triad of one-forms
obeying 
\[
\mathbf{d}{\mbox{\boldmath${\omega}$}}^{a}=-\frac{1}{2}C_{~bc}^{a}{%
\mbox{\boldmath${\omega}$}}^{b}\wedge {\mbox{\boldmath${\omega}$}}^{c},
\]%
and $C_{~bc}^{a}$ depend only on time and are the structure constants
of the Bianchi group type 
under consideration. The structure constants $C_{~bc}^{a}$ can be split into
a vector part $a_{b}$, and a trace-free part $n^{ab}$ by \cite{EM} 
\[
C_{~bc}^{a}=\varepsilon _{bcd}n^{da}-\delta _{~b}^{a}a_{c}+\delta
_{~c}^{a}a_{b}.
\]%
The matrix $n^{ab}$ is symmetric, and, using the Jacobi identity, $%
a_{b}=(1/2)C_{~ba}^{a}$ is in the kernel of $n^{ab}$ 
\[
n^{ab}a_{b}=0.
\]
For the type VI$_0$ model, $a_c=0$ and $n_{ab}$ has two non-zero
eigenvalues with opposite sign. This implies that we can choose  
 a frame such that the structure constants can be written 
\beq
n_{ab}=\begin{bmatrix}
0 &0 & 0 \\
0 & \bar{n} & n \\
0 & n & \bar{n} 
\end{bmatrix}, \qquad a_b=0.
\eeq
Furthermore, the type VI$_0$ has $\bar{n}^2<n^2$.~\footnote{Note that
the Bianchi type II is the limit where $\bar{n}^2=n^2$, and that
Bianchi type VII$_0$ has $\bar{n}^2>n^2$.} The equations
of motion can now be written down. 

Following \cite{DynSys} we introduce expansion-normalised variables to
write the system as an autonomous system of differential equations. In
the notation of ref. \cite{vanElst} we define
\beq
\Sigma_{ab}=\begin{bmatrix} 
-2\Sigma_+ & \sqrt{3}\Sigma_{12} & \sqrt{3}\Sigma_{13} \\
\sqrt{3}\Sigma_{12} & \Sigma_++\sqrt{3}\Sigma_- & \sqrt{3}\Sigma_{23}
\\
 \sqrt{3}\Sigma_{13} & \sqrt{3}\Sigma_{23} & \Sigma_+-\sqrt{3}\Sigma_-
\end{bmatrix}, \nonumber \\  
N_{22}=N_{33}=\sqrt{3}\bar{N},\quad N_{23}=N_{32}=\sqrt{3}N.
\eeq
We also introduce the three-velocity $V$ by
\beq
\sinh\beta =\frac{V}{\sqrt{1-V^2}},\quad 0\leq V<1. 
\eeq

For the expansion-normalised variables the equations of motion are:
\beq
\Sigma_+'&=&
(q-2)\Sigma_++{3}(\Sigma_{12}^2+\Sigma^2_{13})-2N^2 \nonumber \\
&& +\frac{1}{2}\left[4N\Sigma_-v_1+(N\Sigma_{12}+\bar{N}\Sigma_{13})v_2-(N\Sigma_{13}+\bar{N}\Sigma_{12})v_3\right]
\\
\Sigma_-'&=&
(q-2)\Sigma_-+\sqrt{3}(\Sigma_{12}^2-\Sigma_{13}^2)-2R_1\Sigma_{23}\nonumber
\\ &&
+\frac{\sqrt{3}}{2}\left[(N\Sigma_{12}+\bar{N}\Sigma_{13})v_2+(N\Sigma_{13}+\bar{N}\Sigma_{12})v_3\right]
\\
\Sigma'_{12}&=&
\left(q-2-3\Sigma_+-\sqrt{3}\Sigma_-+\sqrt{3}Nv_1\right)\Sigma_{12}
\nonumber \\
&& -\left(R_1+\sqrt{3}\Sigma_{23}-\sqrt{3}\bar{N}v_1\right)\Sigma_{13}
\\
\Sigma'_{13}&=&\left(q-2-3\Sigma_++\sqrt{3}\Sigma_--\sqrt{3}Nv_1\right)\Sigma_{13}
\nonumber \\
&& -\left(-R_1+\sqrt{3}\Sigma_{23}+\sqrt{3}\bar{N}v_1\right)\Sigma_{12} \\
\Sigma'_{23}&=&(q-2)\Sigma_{23}-2\sqrt{3}\bar{N}N+2R_1\Sigma_-+2\sqrt{3}\Sigma_{12}\Sigma_{13}\nonumber
\\
&& -\sqrt{3}v_2(N\Sigma_{13}+\bar{N}\Sigma_{12})\\
N'&=& \left(q+2\Sigma_+\right)N+2\sqrt{3}\Sigma_{23}\bar{N}\\ 
\bar{N}'&=& \left(q+2\Sigma_+\right)\bar{N}+2\sqrt{3}\Sigma_{23}N\\ 
\Omega'&=& \frac{\Omega}{1+(\gamma-1)V^2}\Big\{2q-(3\gamma-2)
 +\left[2q(\gamma-1)-(2-\gamma)-\gamma\mathcal{S}\right]V^2\Big\}
 \quad \\
V'&=&
\frac{V(1-V^2)}{1-(\gamma-1)V^2}\left[(3\gamma-4)-\mathcal{S}\right]
\eeq
where 
\beq
q&=& 2\Sigma^2+\frac
12\frac{(3\gamma-2)+(2-\gamma)V^2}{1+(\gamma-1)V^2}\Omega\nonumber \\
\Sigma^2 &=& \Sigma_+^2+\Sigma_-^2+\Sigma_{12}^2+ \Sigma_{13}^2+\Sigma_{23}^2\nonumber \\
\mathcal{S} &=& \Sigma_{ab}c^ac^b, \quad c^ac_{a}=1, \quad v^a=Vc^a \nonumber \\
 V^2 &=& v_1^2+v_2^2+v_3^2.
\eeq
These variables are subject to the constraints
\beq
1&=& \Sigma^2+N^2+\Omega \label{const:H}\\
0 &=& 2\Sigma_-N\left[1+(\gamma-1)V^2\right]+\gamma\Omega v_1 \label{const:v1}\\
0 &=&
-\left(\Sigma_{12}N+\Sigma_{13}\bar{N}\right)\left[1+(\gamma-1)V^2\right]+\gamma\Omega
v_2 \label{const:v2}\\
0 &=&
\left(\Sigma_{13}N+\Sigma_{12}\bar{N}\right)\left[1+(\gamma-1)V^2\right]+\gamma\Omega
v_3 \label{const:v3} \\
0&=& NR_1-\sqrt{3}\Sigma_-\bar{N} .\label{const:R}
\eeq
The parameter $\gamma$ will be assumed to be in the interval $\gamma\in
( 0,2)$. 

Note that $R_1$, which is the component of the rotation tensor describing rotations with respect to the axis ${\bf e}_1$, is implicitly defined via eq.(\ref{const:R}). For the type VI${_0}$ model we will in practice solve this equation by introducing the parameter $\lambda$ instead of $\bar{N}$, by $\bar{N}=\lambda N$. For the type VI$_0$ model, this parameter is  bounded  by $\lambda^2\leq 1$.

\subsection{The state space}
The constraint (\ref{const:H}) implies that
$\Sigma_{\pm},~\Sigma_{12},~\Sigma_{13},~\Sigma_{23},~N$ and $\Omega$
 are all bounded. Combining the equations for $N$ and $\bar{N}$ we get
the equations 
\beq
\left(N\pm \bar{N}\right)'&=& \left(q+2\Sigma_+\pm
2\sqrt{3}\Sigma_{23}\right)\left(N\pm \bar{N}\right). 
\eeq
Thus if the initial data have $\bar{N}^2<N^2$, then this will hold for all
times. Hence, for type VI$_0$, $\bar{N}$ will be bounded
as well. The invariant subspaces $N\pm\bar{N}=0$ correspond to Bianchi
type II universes. Note that, given the set
$(\Sigma_{\pm},~\Sigma_{12},~\Sigma_{13},~\Sigma_{23},~N,~\bar{N})$ we
can determine $\Omega$ from eq. (\ref{const:H}), and $v^1$, $v^2$ and
$v^3$ from eqs. (\ref{const:v1}), (\ref{const:v2}) and
(\ref{const:v3}), respectively. We also require $0\leq V\leq 1$
for physical reasons (we are not allowing superluminal velocities). This implies the bounds
\beq
\Sigma_+^2+\Sigma_-^2+\Sigma_{12}^2+ \Sigma_{13}^2+\Sigma_{23}^2+N^2
&\leq & 1 \nonumber \\
\bar{N}^2 & \leq  & N^2 \nonumber \\
4\Sigma_-^2N^2+\left(\Sigma_{12}N+\Sigma_{13}\bar{N}\right)^2+\left(\Sigma_{13}N+\Sigma_{12}\bar{N}\right)^2
&\leq& 1.
\label{bounds}\eeq
However, as can be shown, the last of these inequalities is redundant
due to the fact that the first two imply the third (see Appendix \ref{app:proof}). 
Hence, the state space can be considered a subspace of a compact region in
$\mb{R}^7$. 

There are also some discrete symmetries under which this system is
invariant. These are (of course, also compositions of these maps are
symmetries) 
\beq
& \phi_1:& \left(\Sigma_-,\Sigma_{12},\Sigma_{13}\right)\longmapsto 
\left(-\Sigma_-,-\Sigma_{13},-\Sigma_{12}\right) \\
&\phi_2:& \left(\Sigma_{12},\Sigma_{23},\bar{N}\right)\longmapsto 
\left(-\Sigma_{12},-\Sigma_{23},-\bar{N}\right) \\
&\phi_3:& \left(\Sigma_{13},\Sigma_{23},\bar{N}\right)\longmapsto 
\left(-\Sigma_{13},-\Sigma_{23},-\bar{N}\right).
\eeq
To understand the physical importance of these maps we can apply them to
the vector $v^a$: 
\beq
& \phi_1:& \left(v_1,v_2,v_3\right)\longmapsto 
\left(-v_1,v_3,v_2\right) \nonumber \\
&\phi_2:& \left(v_1,v_2,v_3\right)\longmapsto 
\left(v_1,-v_2,v_3\right) \nonumber \\
&\phi_3: &\left(v_1,v_2,v_3\right)\longmapsto 
\left(v_1,v_2,-v_3\right).\nonumber 
\eeq
Hence, these discrete symmetries correspond to different permutation
and inversions of the axes. These symmetries are intrinsic to the type
VI$_0$ geometry. 

\subsection{Invariant subspaces}
\label{sect:inv}
The invariant subspaces play an important role in the evolution of
Bianchi type VI$_0$  universes. They are important to study
since they are invariant under the evolution of the system; i.e. if
a point $p$ lie in one invariant subspace, then also the maximal extended
evolution of $p$ will lie entirely inside it. 

Some of the physically interesting invariant subspaces are as follows
(we will also assume that the boundaries are included):
\begin{enumerate}
\item{} $T(VI_0)$: The full state space of tilted type VI$_0$.
\item{} $F(VI_0)$: The set of fixed points of the map $\phi_1$. Given by
$\Sigma_-=0$, $\Sigma_{12}=-\Sigma_{13}$.  
\item{} $T_2(VI_0)$: A Bianchi type VI$_0$ with a two-component tilted
fluid. It is the set of fixed points of $\phi_2$ (or $\phi_3$). Given by $\Sigma_{12}=\Sigma_{23}=\bar{N}=0$ (or $\Sigma_{13}=\Sigma_{23}=\bar{N}=0$). 
\item{} $T_1(VI_0)$: Bianchi type VI$_0$ with a one-component tilted
fluid. Given by $\Sigma_{12}=\Sigma_{13}=0$. This is the
fixed-point-set of $\phi_2\circ \phi_3$. 
\item{} $B(VI_0)$: Non-tilted Bianchi type VI$_0$. Given by
$\Sigma_-=\Sigma_{12}=\Sigma_{13}=V=0$.
\item{} $T^{\pm}(II)$: The tilted type II boundary. Given by
$N=\pm \bar{N}$. Note that these two subspaces can be mapped onto
each other using $\phi_2$ or $\phi_3$. Because of this, we will in most
cases not differentiate between these two invariant subspaces. 
\item{} $B(II)$: Non-tilted type II. Given by
$N^2=\bar{N}^2$ and $V=0$. 
\item{} $B(I)$: Bianchi type I universes. Given by
$N=\bar{N}=V=0$. 
\item{} $\partial_{V} B(I)$: The Bianchi type I vacuum boundary. Given by
$N=\bar{N}=V=\Omega=0$. 
\end{enumerate}
\begin{table}
\centering
\begin{tabular}{|c|c||c|c|}
\hline
Subspace & Dim & Subspace & Dim  \\ \hline\hline
$T(VI_0)$ & 7 & &  \\ 
$F(VI_0)$ & 5  & $T^{\pm}(II)$ & 6$^*$\\ 
$T_2(VI_0)$ & 4 & $B(II)$ & 4$^*$ \\
$T_1(VI_0)$ & 5 & $B(I)$ & 5$^*$ \\
$B(VI_0)$ & 4 & $\partial_VB(I)$ & 4$^*$  \\ 
\hline
\end{tabular}
\caption{The dimension of the state space for different
invariant subspaces. The asterisk indicates that the true number of
physical degrees of freedom should be reduced  due to unphysical gauge freedoms.}
\end{table}

\section{Equilibrium points of  importance for the late-time
  behaviour}
\label{sect3}
The system of equations has numerous equilibrium points which
correspond to fixed points in the evolution of the system. That is,
if the state variables are written as a vector ${\bf X}$, and we write
the system of equations as
\beq
{\bf X}'={\bf F}({\bf X}),
\eeq 
then a point ${\bf X}_0$ is an equilibrium point if 
\[ {\bf F}({\bf X}_0)=0. \]
These equilibrium points usually play a particular role in the
evolution of the system. Not only are they exact solutions, but some
of these equilibrium points may act as 
past or future attractors for the set of solutions. Hence, for
example, at late times these equilibrium points may be a good
approximation of more general solutions. 

Some of the physically interesting equilibrium
points for the future evolution are given below.  If some
equilibrium points are related via one of the discrete symmetries, 
permutations, or other gauge symmetries,
we only give one of them because the solutions they represent  are equivalent. We also indicate
the smallest invariant subspace mentioned in section \ref{sect:inv} to which
they belong.

\subsection{Non-tilted}
\begin{enumerate}
\item{}$\mathcal{I}(I)$: FRW \\
$\Sigma^2=N=\bar{N}=V=0$, $\Omega=1$, $q=\frac 12(3\gamma-2)$.\\
Invariant subspace: $B(I)$.

\item{}{$\mathcal{CS}(II)$:} Collins-Stewart type II ($2/3<\gamma<2$) \\
$\Sigma_-=\Sigma_{12}=\Sigma_{13}=V=0$,
$\Sigma_+=-\frac{1}{16}(3\gamma-2)$, $\bar{N}=N$ \\
$\Sigma_{23}=-\frac{\sqrt{3}}{16}(3\gamma-2)$, $N^2=\frac
3{64}(3\gamma-2)(2-\gamma)$,\\
$\Omega=\frac 3{16}(6-\gamma)$, $q=\frac 12(3\gamma-2)$.\\
Invariant subspace: $B(II)$
\item{}{$\mathcal{C}(VI_0)$:} Collins VI$_0$ ($2/3<\gamma<2$) \\
$\Sigma_-=\Sigma_{12}=\Sigma_{13}=\Sigma_{23}=\bar{N}=V=0$, $\Sigma_+=-\frac 14(3\gamma-2)$, \\
$N^2=\frac 3{16}(3\gamma-2)(2-\gamma)$, $\Omega=\frac 34(2-\gamma)$,
$q=\frac 12(3\gamma-2)$ \\
Invariant subspace: $B(VI_0)$
\end{enumerate}
\subsection{Intermediately tilted} 
\begin{enumerate}
\item{}$\mathcal{H}(II)$: Hewitt's tilted type II ($10/7<\gamma <2$)
  \cite{HewittII} \\
$\Sigma_-=0$, $\Sigma_+=\frac 18(9\gamma-14)$,
$\Sigma_{23}=-\frac{\sqrt{3}}8(5\gamma-6)$,\\
$\Sigma_{12}=\Sigma_{13}=-\frac{\sqrt{6}}{8}\sqrt{\frac{(2-\gamma)(11\gamma-10)(7\gamma-10)}{17\gamma-18}}$,
$\bar{N}=N$, \\ 
$N^2=\frac{3(2-\gamma)(5\gamma-4)(3\gamma-4)}{4(17\gamma-18)}$,
$V^2=\frac{(3\gamma-4)(7\gamma-10)}{(11\gamma-10)(5\gamma-4)}$, \\
$\Omega=\frac{3(2-\gamma)}{4(17\gamma-18)}(21\gamma^2-24\gamma+4)$,
$q=\frac 12(3\gamma-2)$.\\
Invariant subspace: $T^+(II)$
\item{} $\mathcal{L}(II)$: Type II line bifurcation ($\gamma=14/9$)
  \cite{HBWII} \\
$\Sigma_{\pm}=0$, $\Sigma_{23}=-\frac{2\sqrt{3}}{9}$, $\bar{N}=N$, $N^2=\frac{1}{171}(2b+1)(17-8b)$\\
$\Sigma_{13}+\Sigma_{12}=-\frac 23\sqrt{\frac{2}{57}(4b+1)(8-3b)}$,
$\Sigma_{13}-\Sigma_{12}=-\frac {2\sqrt{2}}{3\sqrt{3}}b$, \\
$V^2=\frac{3(4b+1)(2b+1)}{(17-8b)(8-3b)}$, $\Omega=\frac
2{171}(16b^2-45b+59)$, \\
$0<b<1$, $q=\frac 43$.\\
Invariant subspace: $T^+(II)$
\item{} $\mathcal{R}(VI_0)$: Rosquist-Jantzen ($6/5< \gamma < 3/2$) \cite{RJ}\\
$\Sigma_{13}=\Sigma_{23}=\bar{N}=0$, $\Sigma_+=-\frac 14(3\gamma-2)$, $q=\frac 12(3\gamma-2)$,
\beq
c_1 &= &\frac
12\left\{\frac{(5\gamma-6)\left[9\gamma^2-13\gamma+6+3(2-\gamma)s\right]}{\gamma(9\gamma-10)}\right\}^{\frac
12}\nonumber \\
c_2\nonumber &=&\frac
12\left\{\frac{(2-\gamma)\left[45\gamma^2-65\gamma+18-3(5\gamma-6)s\right]}{\gamma(9\gamma-10)}\right\}^{\frac
12}\nonumber \\ c_3 &=& 0\nonumber 
\eeq
where $s=\sqrt{(\gamma-1)(9\gamma-1)}$. 
\[ \Sigma_-=-\frac 12\delta c_1,\quad \Sigma_{12}=\delta c_2\] 
where
$\delta=\sqrt{3}[(3\gamma-2)-2(2-\gamma)c_2^2]/(6c_1c_2^2)$.
\beq
N^2&=&\frac{(5\gamma-6)^2-c_1^2(3\gamma-4)}{12c_1^2c_2^2} \left[11\gamma-6-(3\gamma+2)c_1^2\right]\nonumber\\
 V&=&\lambda/N, \quad \Omega=1-\Sigma-N^2, \nonumber
\eeq
where $N<0$, and 
\[
\lambda=\frac{1}{2\sqrt{3}}(2-\gamma)\frac{c_1}{c_2^2}-\frac{1}{\sqrt{3}}(5\gamma-6)\frac{1}{c_1}.\]
Invariant subspace: $T_2(VI_0)$
\item{} $\mathcal{L}(VI_0)$: Type VI$_0$ line bifurcation
($\gamma=6/5$)\footnote{These
  solutions seem to be found first by Apostolopoulos
  \cite{Apo}. However, he writes the solutions implicitly as solutions
to a cubic. As can seen, we managed to write them explicitly
down in terms of one free variable which  makes it easier to interpret
the solutions.} \\
$\Sigma_-=\Sigma_{23}=0$, $\Sigma_+=-\frac 25$, $c_1=0$, $q=\frac 45$, \\
$\bar{N}=\lambda N$,
$N=\frac{\sqrt{6}}{5}\sqrt{\frac{5V^2+1}{2V^2+3\lambda+1}}$, \\
$\Sigma_{12}=-\Sigma_{13}=\frac
1{10}\sqrt{\frac{18V^2-6\lambda(5V^2+4)}{2V^2+3\lambda+1}}$,
$\Omega=\frac 3{25}\frac{(V^2+5)(1+5\lambda)}{2V^2+3\lambda+1}$,
\[ V^2=\frac{15+211\lambda+397\lambda^2+25\lambda^3\pm
3(5\lambda+1)\sqrt{s}}{10(3-5\lambda)(1-\lambda)^2} \]
where 
\beq
s =&&
\left[(\lambda+24-9\sqrt{5})^2-16(61-27\sqrt{5})\right]\nonumber \\
&\times & \left[(\lambda+24+9\sqrt{5})^2-16(61+27\sqrt{5})\right]
\nonumber \\
&&-24-9\sqrt{5}+4\sqrt{61+27\sqrt{5}}\leq \lambda < 0. \nonumber 
\eeq
Invariant subspace: $F(VI_0)$
\end{enumerate}

\subsection{Extremely tilted}
\begin{enumerate}
\item{} $\mathcal{E}(II)$: Extremely tilted type II ($0<\gamma<2$)\\
$\Sigma_{\pm}=0$, $\Sigma_{23}=-\frac{2\sqrt{3}}{9}$,
$\Sigma_{13}+\Sigma_{12}=-\frac{10}{3}\sqrt{\frac{2}{57}}$, \\
$\Sigma_{13}-\Sigma_{12}=-\frac{2}{3}\sqrt{\frac{2}{3}}$,
$\bar{N}=N=\sqrt{\frac{3}{19}}$, $V=1$, $\Omega=\frac{20}{57}$,
$q=\frac 43$.\\
Invariant subspace: $T^+(II)$
\item{} $\mathcal{E}_1(VI_0)$: Extremely tilted type VI$_0$
($0<\gamma<2$)\\
$V=1$, $\Sigma_{13}=\Sigma_{23}=\bar{N}=0$, $\Sigma_+=-\frac 58$, $\Sigma_-=\frac {\sqrt{3}}{8}$, 
$\Sigma_{12}=-\frac 14$, 
$N=-\frac 12$, $\Omega=\frac 14$,  $q=\frac{5}{4}$.\\
Invariant subspace: $T_2(VI_0)$
\item{} $\mathcal{E}_2(VI_0)$: Extremely tilted type VI$_0$ 
($0<\gamma<2$)\\
$V=1$, $\Sigma_-=\Sigma_{23}=\bar{N}=0$, $\Sigma_+=-\frac 25$, \\
$\Sigma_{12}=-\Sigma_{13}=\frac {\sqrt{6}}{10}$, 
$N=\frac {2\sqrt{3}}5$, $\Omega=\frac{6}{25}$,  $q=\frac{4}{5}$.\\
Invariant subspace: $F(VI_0)$
\end{enumerate}

\begin{figure}[tbp]
\centering 
\epsfig{figure=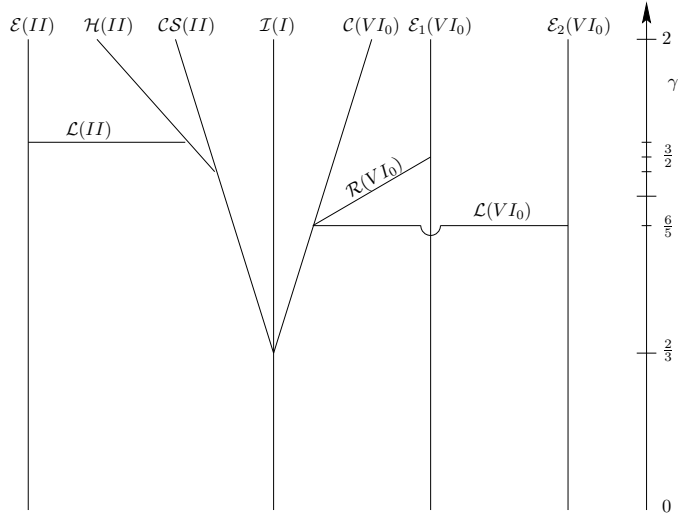, width=9cm}
\caption{The web of future stable equilibrium points. }
\label{web}
\end{figure}

\section{The late-time asymptotic behaviour}
\label{sect4}
An important key to understanding the late-time behaviour of the
dynamical system is to consider the future stable equilibrium
points. Assume that ${\bf X}_0$ is an equilibrium point; i.e. ${\bf
  F}({\bf X}_0)=0$. Then we can write the equations of motion as 
\beq
\delta{\bf X}'={\sf A}\delta{\bf X}+\mathcal{O}(\delta{\bf X}^2).
\eeq 
The local stability of the equilibrium point, ${\bf X}_0$, depends on the
eigenvalues of ${\sf A}$; if all eigenvalues have negative real parts,
then  ${\bf X}_0$ is a future attractor. Every eigenvalue with a
positive real part signals an unstable mode. 

The (local) future stable equilibrium points are for the different invariant
subspaces given in Table 
\ref{stable} and the various connections between  them are illustrated in
Fig.\ref{web}. The eigenvalues of the linearised system are discussed
in Appendix \ref{app:eigen}. 
\begin{table}
\centering
\begin{tabular}{|c|c|c|}
\hline
Space & Matter & Stable point \\ \hline\hline
$B(I)$ & $0<\gamma< 2$ & ${\mathcal I}(I)$ \\ \hline
$B(II)$ &  $0<\gamma< 2/3$ & ${\mathcal I}(I)$ \\
 &  $2/3 <\gamma< 2$ & ${\mathcal{CS}}(II)$ \\  \hline
$T^{\pm}(II)$ &  $0<\gamma< 2/3$ & ${\mathcal I}(I)$ \\
 &  $2/3 <\gamma< 10/7$ & ${\mathcal{CS}}(II)$ \\ 
 &  $10/7<\gamma< 14/9$ & ${\mathcal H}(II)$ \\
 &  $\gamma= 14/9$ & ${\mathcal L}(II)$ \\
&  $14/9<\gamma<2$ & ${\mathcal E}(II)$ \\ \hline
$B(VI_0)$ & $0<\gamma< 2/3$ & ${\mathcal I}(I)$ \\
 &  $2/3 <\gamma< 2$ & ${\mathcal C}(VI_0)$ \\ \hline
$T_1(VI_0)$ &  $0<\gamma< 2/3$ & ${\mathcal I}(I)$ \\
 &  $2/3 <\gamma< 2$ & ${\mathcal C}(VI_0)$ \\ \hline
$T_2(VI_0)$ &  $0<\gamma< 2/3$ & ${\mathcal I}(I)$ \\
 &  $2/3 <\gamma< 6/5$ & ${\mathcal C}(VI_0)$ \\ 
 &  $6/5 <\gamma< 3/2$ & ${\mathcal R}(VI_0)$ \\
 &  $3/2 <\gamma< 2$ & ${\mathcal E}_1(VI_0)$ \\ \hline
$F(VI_0)$ &  $0<\gamma< 2/3$ & ${\mathcal I}(I)$ \\
 &  $2/3 <\gamma< 6/5$ & ${\mathcal C}(VI_0)$ \\
 &  $\gamma=6/5 $ & ${\mathcal L}(VI_0)$  \\ 
 &  $6/5 <\gamma< 2$ & ${\mathcal E}_2(VI_0)$  \\ \hline
$T(VI_0)$ &  $0<\gamma< 2/3$ & ${\mathcal I}(I)$ \\
 &  $2/3 <\gamma< 6/5$ & ${\mathcal C}(VI_0)$ \\ 
 &  $\gamma= 6/5$ & ${\mathcal L}(VI_0)$  \\
 &  $6/5 <\gamma< 2$ & ${\mathcal E}_2(VI_0)$  \\ 
\hline
\end{tabular}
\caption{The future stable equilibrium points for various invariant (sub)spaces.}\label{stable}
\end{table}

For the non-tilted subspaces $B(I)$, $B(II)$, and $B(VI_0)$, it is proven that
generic solutions approach the respective local future attractors \cite{DynSys}. In
the tilted case we can only partially prove that the local attractors are
also global attractors. The subspace $T^{\pm}(II)$ was discussed in
\cite{HBWII}. We will use these results in the following way. Assume we have a subspace, $\mathcal{S}$ (given by $y=0$, say), for which a function $Z$ is strictly monotonically decreasing; i.e. 
\[\left.Z'\right|_{y=0}=\alpha\left.Z\right|_{y=0}, \quad \alpha \leq \epsilon <0. \]
Away from $\mathcal{S}$, we can write $Z'=\alpha(y)Z$ where $\alpha(0)\leq \epsilon <0$. The function $\alpha(y)$ is continuous in $y$; hence, there will exist a $\delta>0$ such that 
\beq
\alpha(y)<0, \quad \text{for} \quad |y|<\delta. 
\eeq
So suppose the solution curves have the property $y\rightarrow 0$, then after sufficently long time the function $Z$ will be monotonic along the solution curves. The late time analysis can therefore to some extent be extracted from the monotonic functions in $\mathcal{S}$. 

In the following,
$\omega(p)$ will 
mean the future 
asymptotic set; i.e. for a solution curve $c(\tau)$ having $c(0)=p$,
then $\omega(p)=\lim_{\tau\rightarrow\infty}c(\tau)$. 
\begin{thm}[No-hair] 
For $\gamma$ obeying $0<\gamma<2/3$, any $p\in T(VI_0)$ for which
$V < 1$, and $\Omega>0$ has
$\omega(p)=\mathcal{I}(I)$. 
\end{thm}
\begin{proof}
Part of the proof relies on the observation that the function
 $\mathcal{S}=\Sigma_{ab}c^ac^b$ obeys the bound $|\mathcal{S}|\leq
 2\Sigma\leq 2$ (see Appendix \ref{app:proof}). From the equation for
 $V$ we thus have
\[ 
V'\leq (3\gamma-2)\frac{V(1-V^2)}{1-(\gamma-1)V^2}.
\] 
Hence, if $0<\gamma<2/3$, $V$ will be monotonically decreasing.
 
There is also an other monotonically increasing function, namely
\beq
\left(\beta\Omega\right)'&=&\left[2q-(3\gamma-2)\right]\left(\beta\Omega\right),\\
\beta &\equiv & \frac{(1-V^2)^{\frac 12(2-\gamma)}}{1+(\gamma-1)V^2}.
\eeq
Note that 
\[
2q-(3\gamma-2)=3(2-\gamma)\Sigma^2+(2-3\gamma)N^2+\frac{\gamma(4-3\gamma)V^2}{1+(\gamma-1)V^2}\Omega.
\] 
Hence, it follows that for $0<\gamma<2/3$,  $\lim_{\tau \ra \infty}\Omega= 1$, and
$\lim_{\tau \ra \infty}V= 0$. 
\end{proof}

\begin{thm}[Global attractors for $T_1(VI_0)$]
For $2/3<\gamma<4/3$, any $p\in T_1(VI_0)$ with $\Omega> 0$,
$N^2>\bar{N}^2$ and $V<1$, has $\omega(p)=\mathcal{C}$. 
\end{thm}
\begin{proof}
We note that in the subspace $T_1(VI_0)$ we have 
\beq
\Sigma_+=(q-2)\Sigma_+-2N^2-\frac{\gamma\Omega v_1^2}{1+(\gamma-1)V^2}.
\eeq
Since $q\leq 2$ (which can easily be checked) there exist a $\tau_1$ such that $\Sigma_+\leq 0$ for all $\tau>\tau_1$. Furthermore, in $T_1(VI_0)$ we also have 
\beq
V'=\frac{V(1-V^2)}{1+(\gamma-1)V^2}\left[(3\gamma-4)+2\Sigma_+\right]. 
\eeq
Hence, for $\gamma < 4/3$, $V$ will be monotonically decreasing for $\tau>\tau_1$, and $\lim_{\tau\rightarrow\infty}V=0$. This means that the solutions will approach the non-tilted subspace $B(VI_0)$ in the limit $\tau \rightarrow \infty$. Thus after sufficiently long time, the monotone functions in $B(VI_0)$ (given in \cite{DynSys}) will also be monotone along orbits in $T(VI_0)$. 
The point $\mathcal{C}$ is an isolated equilibrium point, thus we can use the monotone functions which implies $\omega(p)=\mathcal{C}$.
\end{proof}
There are strong reasons to believe that $\mathcal{C}$ is a global attractor in $T_1(VI_0)$ for all $0<\gamma <2$. As evidence for this is the existence of a monotonic function in $T_1(VI_0)$. Define $\sigma=\Sigma_-^2+\Sigma_{23}^2$, which in $T_1(VI_0)$ has
the evolution equation
\beq
\sigma'=2(q-2)\sigma-4\sqrt{3}N\bar{N}\Sigma_{23}.
\eeq
In $T_1(VI_0)$ we have the monotonically increasing function:
\beq
Z_1\equiv \frac{N^2+\sigma}{\bar{N}^2+\sigma},\quad
Z_1'=\frac{4\sigma(N^2-\bar{N}^2)(\Sigma_++1)}{(N^2+\sigma)(\bar{N}^2+\sigma)}Z_1. 
\eeq 
As $\tau\rightarrow\infty$ there are thus three possibilities: 
\[ \sigma, \quad \Sigma_++1, \quad  \text{ or } N^2-\bar{N}^2 \rightarrow 0.\] 
If $\sigma\rightarrow 0$, the non-tilted
analysis can again be applied. The case $\Sigma_+=-1$ is unstable in the future so this cannot happen for general solutions. If $N^2-\bar{N}^2\rightarrow 0$, the 
tilted analysis of type II can be applied. All the late-time
asymptotes of the type II case have an unstable 
direction into the interior of $T_1(VI_0)$ (which exactly corresponds
to $N^2-\bar{N}^2$). Unfortunately, for the tilted type II model the local attractors have not been rigorously proven to be global attractors \cite{HBWII}. 

Apart from these two theorems we have not been able to show any global
late time attractors for the various (sub)spaces. However, numerical
analysis seems to imply that the local attractors also are
global attractors. 

\section{The initial singular regime}
\label{sect5}
Let us consider the initial singular regime; i.e. where $\tau \ra
-\infty$. This case is a lot more subtle than at late times due to the
oscillatory behaviour of the system of equations as one approach  $\tau \ra
-\infty$. In fact, the tilted type II seems to have an initial
oscillatory regime \cite{HBWII}, and hence, 
one would expect a similar -- if not more complex -- behaviour for the tilted type VI$_0$ model. 

In the following we should emphasise that we consider a  non-stiff fluid.  For a stiff fluid ($\gamma=2$) it has been pointed out that the Bianchi models allow for a stable past attractor which would remove this chaotic behaviour into the past \cite{barrow78,coley}.

In the study of the initial singular regime it is useful to introduce
the variable $\lambda$, instead of $\bar{N}$, as
\beq
\bar{N}=\lambda N.
\eeq
The equations for $N$ and $\lambda$ are then
\beq
N' &=& (q+2\Sigma_++2\sqrt{3}\lambda\Sigma_{23})N \\
\lambda' &=& 2\sqrt{3}\Sigma_{23}(1-\lambda^2).
\eeq
In this case, the invariant subspaces $\lambda=\pm 1$ correspond to
$T(II)^{\pm}$. 
In this case, there are two Kasner circles, which differ by an
orientation of frame. They are as follows.\footnote{With a slightly
abuse of notation because they are really type I solutions.}
\begin{enumerate}
\item{} $\mathcal{K}^{\pm}(II)$: Kasner ``type II'' ($0<\gamma<2$)\\
$\Sigma_+^2+\Sigma_{23}^2=1$, $\lambda=\pm 1$, $q=2$ \\ 
$\Sigma_-=\Sigma_{13}=\Sigma_{12}=N=\Omega=0$.
\item{} $\mathcal{K}(VI_0)$: Kasner ``type VI$_0$'' ($0<\gamma<2$)\\
$\Sigma_+^2+\Sigma_{-}^2=1$, $\lambda=0$, $q=2$ \\ 
$\Sigma_{23}=\Sigma_{13}=\Sigma_{12}=N=\Omega=0$.
\end{enumerate}

Each of them also have extremely tilted Kasner sets and corresponding
bifurcations. These are discussed in detail in \cite{UEWE}. We will
not discuss these here as they do not change our conclusion
radically. However, bear in mind that these equilibrium points exist
and that there are transitions between them. 

The behaviour near the initial depends on three types of heteroclinic
orbits. These are
\begin{enumerate}
\item{} $\mathcal{T}_{R_i}$: Frame rotations.
\item{}  $\mathcal{T}_{N{\pm}}$: Taub type II vacuum orbits.
\item{} $\mathcal{T}_{\lambda}$: Frame rotations between  $\mathcal{K}^{\pm}(II)$ and
$\mathcal{K}(VI_0)$. 
\end{enumerate} 
Let us consider, for illustration, the Kasner circle
$\mathcal{K}^+(II)$. The analysis for the Kasner circle
$\mathcal{K}^-(II)$ can be obtained from $\mathcal{K}^+(II)$ by using
the map $\phi_2$ (or $\phi_3$). 
\begin{figure}[tbp]
\centering 
\epsfig{figure=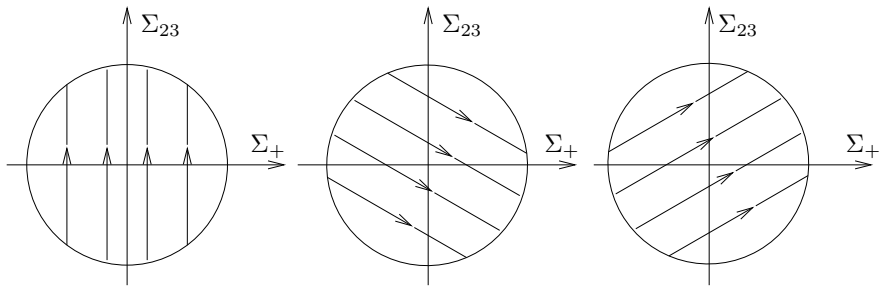, width=12cm}
\caption{The $\mc{K}^+(II)$ frame rotations projected onto the
  $(\Sigma_+,\Sigma_{23})$-plane. Arrows are future-directed.}
\label{framerot}\end{figure}
\begin{figure}[tbp]
\centering 
\epsfig{figure=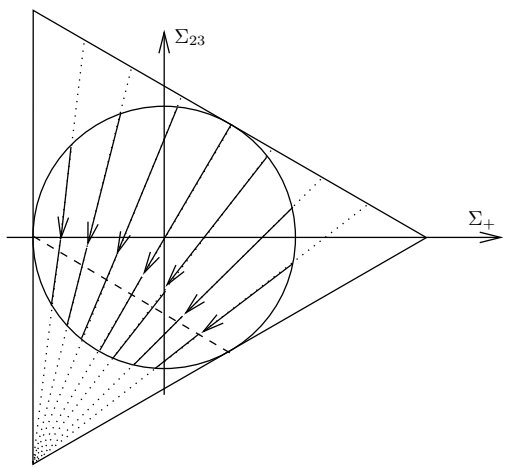, width=7cm}
\caption{Taub orbits, $\mc{T}_{N+}$, projected onto the $(\Sigma_+,\Sigma_{23})$-plane.  Arrows are future-directed.}
\label{Tauborbits}\end{figure}

\subsection*{$\mathcal{K}^+(II)$}
Here, the frame rotations, and Taub type II orbits are given by (when
projected onto the $(\Sigma_+,\Sigma_{23})$-plane):
\begin{enumerate}
\item{} Three frame rotations: \\
$\Sigma_+=C_1$ \\ 
$\Sigma_+\pm\sqrt{3}\Sigma_{23}=C_{\pm}$.
\item{} Taub type II vacuum orbits:\\
$\Sigma_++1=C(\Sigma_{23}+\sqrt{3})$.
\end{enumerate}
All of these orbits maps the Kasner circle onto itself:
$\mathcal{K}^+(II)\mapsto \mathcal{K}^+(II)$. The frame rotations are
illustrated in Fig.\ref{framerot} and the Taub orbits are illustrated
in Fig.\ref{Tauborbits}. 

For the Kasner circle $\mathcal{K}(VI_0)$ there are two frame
rotations of a similar kind as for $\mathcal{K}^{\pm}(II)$; namely the
ones given by $\Sigma_+\pm\sqrt{3}\Sigma_{-}=C_{\pm}$. These orbits
also maps the Kasner circle onto itself: $\mathcal{K}(VI_0)\mapsto
\mathcal{K}(VI_0)$.

In addition to these homoclinic orbits there are homoclinic orbits
between the three types of Kasner circles. They are given by the set of
differential equations
\beq
\lambda'&=& 2\sqrt{3}\Sigma_{23}(1-\lambda^2) \nonumber \\
\Sigma_-' &=& -2\sqrt{3}\lambda\Sigma_-\Sigma_{23} \nonumber \\
\Sigma_{23}'&=& 2\sqrt{3}\lambda\Sigma_-^2 \\
1&=& \Sigma_+^2+\Sigma_-^2+\Sigma_{23}^2, \quad
\Sigma_+=\mathrm{constant}.\nonumber 
\eeq
\begin{figure}[tbp]
\centering 
\epsfig{figure=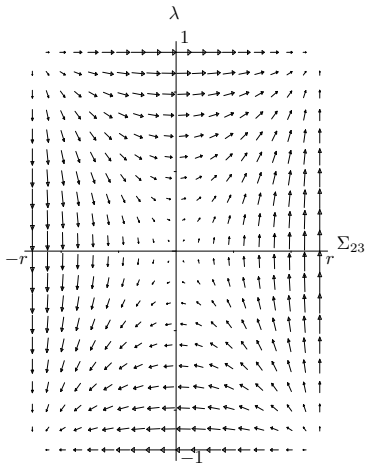, width=5cm}
\caption{Flow diagram for the transitions between the Kasner circles
$\mathcal{K}^{\pm}(II)$ and $\mathcal{K}(VI_0)$ projected onto the $(\Sigma_{23},\lambda)$-plane. Here, the constant
$r$ is given by $r=\sqrt{1-\Sigma_+^2}$. Arrows are future-directed. }
\label{flow}\end{figure}

A flow diagram for this system projected onto the
$(\lambda,\Sigma_{23})$-plane is shown in Fig.\ref{flow}. We can
see that there are homoclinic orbits between  $\mathcal{K}^{\pm}(II)$
and $ \mathcal{K}(VI_0)$. However, these only form a set of measure
zero. Most orbits connect $\mathcal{K}^{\pm}(II)$. Nonetheless, orbits can come arbitrary close to $
\mathcal{K}(VI_0)$ and hence, there might be frame rotations within
the Kasner circle $\mathcal{K}(VI_0)$ before the orbit go back to
$\mathcal{K}^{\pm}(II)$. 

From this fairly simple, but far from complete analysis, we believe
that the initial behaviour of the tilted type VI$_0$ models is fairly
complicated. There are infinite sequences of homoclinic orbits which
consists of frame rotations and Taub vacuum orbits. These orbits are
believed to be chaotic in general and thus we conjecture that \emph{the tilted
type VI$_0$ has a  chaotic behaviour to the past}.

\section{Summary}
\label{sect6}
Here, for the first time, we have analysed the late-time behaviour of
general tilted Bianchi type VI$_0$ universes. Our results are
summarised in Table \ref{stable}. We performed a local analysis and found all the future 
stable equilibrium points for various subclasses of tilted type VI$_0$
models as well as for the general tilted type VI$_0$. In
particular, we confirmed the observation in \cite{BHtilted} on the
existence of new self-similar solutions 
for $\gamma\geq 6/5$. These solutions also proved to be important for
the late-time behaviour; the Rosquist-Jantzen solutions are late-time
attractors for a certain class of models with a two-component tilted
fluid; and the line bifurcation at $\gamma=6/5$ is, in fact, the late-time
attractor for general tilted type VI$_0$ solutions. For  $\gamma>6/5$,
the late-time attractor is an extremely tilted model.\footnote{It
  should be noted that the $\gamma=6/5$ line bifurcation looks
  remarkably similar to the Wainwright $\gamma=10/9$ line bifurcation
  of the exceptional model \cite{Wainwright10-9}. This makes us wonder
if there are unknown line bifurcations along the entire line given by
$\gamma=2(3+\sqrt{-h})/(5+3\sqrt{-h})$ for the more general type
VI$_h$ models \cite{BHtilted}.}

It is interesting to note that the general late-time attactors lie in
the fixed-point-set of $\phi_1$. This means that in general the
solutions are asymptotically  $\phi_1$-symmetric. More specifically,
$\phi_1$-symmetric implies $v_1=0$, and $v_2^2=v_3^2$. 

At early times the analysis suggests chaotic behaviour as $\tau\rightarrow -\infty$. This behaviour seems to be a
generic property of tilted Bianchi models \cite{UEWE}. So this
behaviour was not very surprising, in particular considering the fact
that the tilted type II model -- which is part of the boundary of type
VI$_0$ -- was known  to be chaotic \cite{HBWII}. 

Let us finish off with some comments about future research and how
this work may relate to other Bianchi types. Firstly, the work \cite{BHtilted}
indicates that there may be some similarities between the more general
type VI$_h$ models and the type VI$_0$. Hence, some of the features of
the late-time behaviour found in this work may also appear in the type
VI$_h$ models. As goes for the class VII$_0$ model, which is given by
$\bar{N}^2>N^2$, there is one obvious difference. The general
non-tilted VII$_0$ model is not asymptotically self-similar
\cite{WHU,NHW}. This 
happens because the type VII$_0$ state space is not compact; there is
no upper bound on $\bar{N}$. In the terminology of \cite{BH}, the type
VII$_0$ model is \emph{extremely Weyl dominant} while the type
VI$_0$ model is \emph{Weyl-Ricci balanced} at late times. Hence, we
would expect a fairly different behaviour for the tilted type VII$_0$
model at late times than that found here. Moreover, in this work the extremely tilted invariant sets have not been emphasised. For example, a two-fluid model where one fluid is extremely tilted is if particular interest. Further investigations in this case is required. 

Nonetheless, there are  quite a few unanswered questions regarding 
Bianchi models with a perfect fluid. This work has answered some of them. Hopefully, future work will answer more of them. 

\section*{Acknowledgments}
The author would like to thank J.D. Barrow, A.A. Coley and S.T.C. Siklos for discussions related to this work. 
This work was funded by the Research Council of Norway and an Isaac
Newton Studentship. 

\appendix

\section{Some simple proofs}
\label{app:proof}
\subsection*{On bound (\ref{bounds})}
We start with assuming that the first two bounds in eq. (\ref{bounds})
hold. Then, using the Schwarz inequality, we have
\beq
&& 4\Sigma_-^2N^2+\left(\Sigma_{12}N+\Sigma_{13}\bar{N}\right)^2+\left(\Sigma_{13}N+\Sigma_{12}\bar{N}\right)^2\nonumber
\\ 
&\leq & 
4\Sigma_-^2N^2+2\left(\Sigma_{12}^2+\Sigma_{13}^2\right)\left(N^2+\bar{N}^2\right) \nonumber
\\ &\leq & 4\left(\Sigma_-^2+\Sigma_{12}^2+\Sigma_{13}^2\right)N^2.
\label{proof}\eeq
Consider now the maximal value of the function  $F(X,Y)=4X^2Y^2$
inside the unit disc
\[ X^2+Y^2\leq 1. \]
One easily finds that the maximal value is at $X^2=Y^2=1/2$. Thus 
\[ F(X,Y)\leq 1. \]
Hence, by identifying $X^2=N^2$ and
$Y^2=\Sigma_-^2+\Sigma_{12}^2+\Sigma_{13}^2$, we obtain from
eq. (\ref{proof})
\beq
 4\Sigma_-^2N^2+\left(\Sigma_{12}N+\Sigma_{13}\bar{N}\right)^2+\left(\Sigma_{13}N+\Sigma_{12}\bar{N}\right)^2
 \leq 1. 
\eeq
Thus the last inequality in eq. (\ref{bounds}) is redundant. 
\subsection*{Showing $|\mathcal{S}|\leq 2\Sigma\leq 2$}
We will consider the function
\[ |\mathcal{S}|=|\Sigma_{ab}c^ac^b|, \]
where the matrix $\Sigma_{ab}$ is symmetric and trace-free, and
$c^ac_a=1$. This implies that the maximal value of $|\mathcal{S}|$
occurs when $c^a$ is parallel to one of the eigenvectors of $\Sigma_{ab}$. Thus, if 
$\lambda_i$ are the eigenvalues, we have
\beq
 |\Sigma_{ab}c^ac^b|\leq \max(|\lambda_1|,|\lambda_2|,|\lambda_3|).
\label{eq:lambdas}\eeq
The eigenvalues obey the relations
\beq
\Sigma^a_{~a}=\lambda_1+\lambda_2+\lambda_3&= & 0 \nonumber \\
\Sigma_{ab}\Sigma^{ab}=\lambda_1^2+\lambda_2^2+\lambda^2_3 &= &6\Sigma^2.\nonumber
\eeq
These two equations imply that the maximal value of an eigenvalue
occur when $\lambda_1=\pm 2\Sigma$, and
$\lambda_2=\lambda_3=\mp\Sigma$ (or a permutation thereof). Hence,
according to eqs.(\ref{eq:lambdas}) and (\ref{const:H}) we have
\beq
|\mathcal{S}|\leq 2\Sigma\leq 2.
\eeq

\section{Eigenvalues of Equilibrium points}\label{app:eigen}
In this appendix we discuss some of the eigenvalues for the various
equilibrium points. 
\subsection{Non-tilted}
\begin{enumerate}
\item{}$\mathcal{I}(I)$: FRW \\
\[ \lambda_{1,2,3,4,5}=-\frac 32(2-\gamma),\quad \lambda_{6,7}=\frac
12(3\gamma-2). \]
\item{}{$\mathcal{CS}(II)$:} Collins-Stewart type II ($2/3<\gamma<2$) \\
The essential unstable eigenvalue is for all equilibrium points in
$T(II)$:
\[ \lambda_7=-4\sqrt{3}\Sigma_{23}. \]
Hence, since $\Sigma_{23}<0$  this point is unstable. 
\item{}{$\mathcal{C}(VI_0)$:} Collins VI$_0$ ($2/3<\gamma<2$) \\
\beq& \lambda_{1,2}=-\frac
34(2-\gamma)\left(1\pm\sqrt{5\gamma-6}\right),\quad \lambda_{3,4}=-\frac
34(2-\gamma)\left(1\pm\sqrt{\frac{10-13\gamma}{2-\gamma}}\right),\nonumber \\
&\lambda_5=-\frac 32(2-\gamma),\quad \lambda_{6,7}=-\frac 34(6-5\gamma).
\nonumber 
\eeq
Here, $\lambda_{1,2,3,4}$ correspond to the non-tilted case,
$\lambda_{5,6,7}$ are the eigenvalues in the $v_1$, $v_2$ and $v_3$
directions, respectively. 
\end{enumerate}
\subsection{Intermediately tilted} 
\begin{enumerate}
\item{}$\mathcal{H}(II)$: Hewitt's tilted type II ($10/7<\gamma <2$)\\
Unstable due to the eigenvalue
\[ \lambda_7=-4\sqrt{3}\Sigma_{23}. \]
\item{} $\mathcal{L}(II)$: Type II line bifurcation ($\gamma=14/9$) \\
Unstable due to the eigenvalue
\[ \lambda_7=-4\sqrt{3}\Sigma_{23}. \]
\item{} ${\mathcal R}(VI_0)$: Rosquist-Jantzen ($6/5<\gamma<3/2$) \\
Due to the complex character of these solutions, one has to part of
the  stability analysis numerically. Some eigenvalues are
possible to find analytically. The following seems to hold:
\beq  &\mathrm{Re}(\lambda_{1,2,3,4})=-\frac
34(2-\gamma),\nonumber \\ 
&\lambda_5+\lambda_6=-\frac 32(2-\gamma), \qquad
\mathrm{Re}(\lambda_5),\mathrm{Re}(\lambda_6)<0, \nonumber\\
&\lambda_7=\frac 32(5\gamma-6). \nonumber
\eeq
Here, $\lambda_{1,2,5,6}$ correspond to directions along the  invariant subspace $T_2(VI_0)$. 
\item{} $\mathcal{L}(VI_0)$: Type VI$_0$ line bifurcation
($\gamma=6/5$) \\
Again we have to rely on some numerical analysis. However, analytic
combined with numerics seem to indicate that 
\beq &\lambda_1=0, \qquad \mathrm{Re}(\lambda_{2,3,4,5})=-\frac 35,\nonumber \\
&\lambda_6+\lambda_7=-\frac 65, \qquad \mathrm{Re}(\lambda_6),\mathrm{Re}(\lambda_7)<0.
\nonumber 
\eeq
Here, $\lambda_{1,2,3,4,5}$ correspond to directions along the invariant subspace $F(VI_0)$.
\end{enumerate}
\subsection{Extremely tilted}
\begin{enumerate}
\item{} $\mathcal{E}(II)$: Extremely tilted type II ($0<\gamma<2$)\\
Unstable due to the eigenvalue
\[ \lambda_7=-4\sqrt{3}\Sigma_{23}. \]
\item{} $\mathcal{E}_1(VI_0)$: Extremely tilted type VI$_0$
($0<\gamma<2$)\\
\beq 
&\lambda_{1,2}=-\frac 18\left(3\pm i\sqrt{183}\right),\quad
\lambda_{3,4}=-\frac 18\left(3\pm i\sqrt{3}\sqrt{53+8\sqrt{3}}\right)\nonumber \\
&\lambda_5=-\frac 34,\qquad
\lambda_6=-\frac{3(2\gamma-3)}{2-\gamma},\quad \lambda_7=\frac 94
\eeq
Here, $\lambda_{1,2,5,6}$ correspond to directions along the invariant subspace $T_2(VI_0)$. 
\item{} $\mathcal{E}_2(VI_0)$: Extremely tilted type VI$_0$ 
($0<\gamma<2$)\\
\beq & \lambda_1=\frac{6(6-5\gamma)}{5(2-\gamma)},\quad
\lambda_{2,3}=-\frac 35\left(1\pm i\sqrt{19}\right),\nonumber \\
&\lambda_{4,5}=-\frac 35\left(1\pm i\sqrt{11}\right),\quad \lambda_{6,7}=-\frac 35\left(1\pm i\sqrt{14+5\sqrt{2}}\right).
\eeq
Here, $\lambda_{1,2,3,4,5}$ correspond to directions along the invariant subspace $F(VI_0)$.
\end{enumerate}

\end{document}